\documentclass[preprint,nofootinbib]{revtex4-1}
\usepackage{graphicx}
\usepackage{dcolumn}
\usepackage{bm}
\usepackage{amsmath}
\usepackage{amsfonts} 
\usepackage{latexsym}
\usepackage{bbm}
\usepackage{color}
\newcommand{\non}{\nonumber}

\begin{document}


\title{Improved sphaleron decoupling condition and the Higgs coupling constants in the real singlet-extended SM}

\author{Kaori Fuyuto}%
\email{fuyuto@th.phys.nagoya-u.ac.jp}
\author{Eibun Senaha}%
\email{senaha@eken.phys.nagoya-u.ac.jp}
\affiliation{Department of Physics, Nagoya University, Nagoya 464-8602, Japan}

\bigskip

\date{\today}

\begin{abstract}
We improve the sphaleron decoupling condition in the real singlet-extended standard model (SM).
The sphaleron energy is obtained using the finite-temperature one-loop effective potential
with daisy resummation. 
For moderate values of the model parameters, the sphaleron decoupling condition is found to be 
$v_C/T_C>(1.1-1.2)$, 
where $T_C$ denotes a critical temperature and $v_C$ is the corresponding vacuum 
expectation value of the doublet Higgs field at $T_C$.
We also investigate the deviation of the triple Higgs boson coupling from its standard model value
in the region where the improved sphaleron decoupling condition is satisfied.
As a result of the improvement, the deviation of the triple Higgs boson coupling gets more enhanced.
In a typical case, if the Higgs couplings to the gauge bosons/fermions deviate from the SM values 
by about 3 (10)\%, 
the deviation of the triple Higgs boson coupling can be as large as about 16 (50)\%, 
which is about 4 (8)\% larger than that based on the conventional criterion $v_C/T_C>1$.
\end{abstract}

\pacs{Valid PACS appear here}

\maketitle

\section{Introduction}
Obtaining full knowledge of the Higgs sector is one of the primary goals in collider physics.
Since the discovery of the Higgs boson at the Large Hadron Collider (LHC) in 2012 \cite{Aad:2012tfa,Chatrchyan:2012ufa}, 
much attention has been paid to the question of whether the Higgs sector is exactly the same as the one
predicted by the standard model (SM).
Since many new physics models predict an augmented Higgs sector, 
we may see a new physics signal in precision measurements of the Higgs coupling constants,
such as the Higgs couplings to the weak bosons, fermions, and self couplings.

One of the motivations for new physics is the baryon asymmetry of the Universe (BAU).
The baryon-to-photon ratio is determined by the cosmic microwave background (CMB)
and big bang nucleosynthesis (BBN)~\cite{Beringer:1900zz}:
\begin{align}
\eta_{\rm CMB}  &= \frac{n_B}{n_\gamma} = (6.23\pm0.17)\times 10^{-10}, \\
\eta_{\rm BBN}  &= \frac{n_B}{n_\gamma} = (5.1-6.5)\times 10^{-10},~(95\%~{\rm C.L.}).
\end{align}
To explain the observed value, the so-called Sakharov criteria~\cite{Sakharov:1967dj} are needed, i.e.,
(i) baryon number ($B$) violation, (ii) both $C$ and $CP$ violation, and (iii) the departure from thermal equilibrium. 
It is well known that the last condition in the Sakharov criteria, which can be realized 
if the electroweak phase transition (EWPT) is first order,  
is not satisfied for the observed Higgs boson mass, 126 GeV. 
Lattice calculations show that the EWPT in the SM is a smooth crossover~\cite{sm_ewpt}.
In addition to the above problem, $CP$ violation that comes from the Cabibbo-Kobayashi-Maskawa matrix \cite{Cabibbo:1963yz,Kobayashi:1973fv} is not large enough to generate the sufficient BAU~\cite{ewbg_sm_cp}.

Although many working mechanisms for generating the BAU exist in the literature, 
electroweak baryogenesis (EWBG)~\cite{ewbg} is one of the most testable scenarios
since the relevant energy scale is within our reach, i.e., $\mathcal{O}(100)$ GeV.
It is expected that ongoing and upcoming experiments on the Higgs sector and $CP$ violation 
--- such as electric dipole moments (EDMs) of the neutron, atoms and molecules --- 
can entirely verify or falsify the EWBG hypothesis.
Indeed, EWBG in the minimal supersymmetric standard model (MSSM)
is unlikely in light of the LHC data, especially the Higgs signal 
strengths~\cite{MSSM-EWBG_LHCtension}. 
On the other hand, EWBG in other models is not so severely constrained by the current LHC data. 
In any case, to get a definitive conclusion, theoretical uncertainties have to be minimized.
In particular, the improvement of the sphaleron decoupling condition is indispensable.
Conventionally, the following rough criterion is frequently used:
\begin{align}
\frac{v_C}{T_C}\gtrsim1,
\end{align}
where $T_C$ is a critical temperature and $v_C$ is the corresponding Higgs
vacuum expectation value (VEV) at $T_C$.
Note that the right-hand side also inherently depends on the temperature.
To evaluate the right-hand side more precisely,
the sphaleron profile and the nucleation temperature ($T_N$)
have to be determined.
In Ref.~\cite{Funakubo:2009eg}, the sphaleron energy and zero-mode factors of the fluctuations
around the sphaleron were evaluated at $T_N$ using the finite-temperature one-loop effective potential
in the MSSM. It was found that the sphaleron decoupling condition is  $v_N/T_N\gtrsim 1.4$. 
Therefore, the conventional criterion leads to about 40\% overestimated results regarding the successful EWBG region. 
As demonstrated in Ref.~\cite{Funakubo:2009eg}, the viability of EWBG in the MSSM 
was already in jeopardy due to the refined sphaleron decoupling condition,
independent of the recent LHC data.

In this paper, we improve the sphaleron decoupling condition
and revisit the region where the EWPT is strongly first order in the real singlet-extended SM (rSM).
Earlier related work on this subject can be found in Refs.~\cite{EWBG_rSM,Profumo:2007wc,Gonderinger:2009jp,Ashoorioon:2009nf,Espinosa:2011ax,Cline:2012hg}.
To evaluate the sphaleron decoupling condition, we calculate the sphaleron energy
using the finite-temperature one-loop effective potential with daisy resummation.
We first examine the dependencies of the second Higgs mass and the mixing angle between the
two Higgs bosons on the sphaleron energy without including the temperature effects, 
and then scrutinize the temperature effects.
With the same effective potential, we compute $v_C/T_C$ and search for a parameter space 
that satisfies the improved sphaleron decoupling condition.

We also study relationships between the strength of the strong first-order EWPT
and the deviations of the Higgs coupling constants from the SM values.
It is known that in the region where the strong first-order EWPT is driven, a significant deviation of the
triple Higgs boson coupling from the SM prediction can arise in the two-Higgs-doublet model~\cite{Kanemura:2004ch}.
In Ref.~\cite{Kanemura:2004ch}, it was shown that the deviation can be greater than 10$-$20\% depending on the magnitude of the $Z_2$ breaking mass parameter.
The degree to which the deviation can occur actually depends on the sphaleron decoupling condition.
For instance, if the sphaleron decoupling condition gets more severe, the corresponding deviation of the 
triple Higgs boson coupling can be more enhanced. In the analysis of Ref.~\cite{Kanemura:2004ch}, 
the conventional criterion $v_C/T_C>1$ was used.
In this article, however, we study the deviation of the triple Higgs boson coupling
based on the improved sphaleron decoupling condition.

The paper is organized as follows. 
In Sec.~\ref{sec:Model}, we introduce the model and discuss the vacuum structure of this model.
In Sec.~\ref{sec:EWPT}, we present standard formulas for studying the EWPT
and classify the patterns of the EWPT.
The sphaleron decoupling condition is given in Sec.~\ref{sec:sphaleron}. Subsequently,
a typical example is given in order to see the magnitude of the sphaleron energy in this model.
Our main results are presented in Sec.~\ref{sec:numerics}, and 
Sec.~\ref{sec:conclusion} is devoted to conclusions and discussion.

\section{The model}\label{sec:Model}
We consider a minimal extension of the SM that includes a gauge singlet real scalar $S$. 
The most general renormalizable scalar potential is given by
\begin{align}
V_0=&-\mu^2_HH^{\dagger}H+\lambda_H(H^{\dagger}H)^2\nonumber\\
&+\mu_{HS}H^{\dagger}HS+\frac{\lambda_{HS}}{2}H^{\dagger}HS^2\nonumber\\
&+\mu^3_SS+\frac{m^2_S}{2}S^2+\frac{\mu^{\prime}_S}{3}S^3+\frac{\lambda_S}{4}S^4,\label{treepotential}
\end{align} 
where $H$ is the SU(2) doublet Higgs field. 
After two scalar fields $H$ and $S$ get VEVs ($v$ and $v_S$), 
they are cast into the form
\begin{align}
H(x)=
\begin{pmatrix}
G^+(x)\\
\frac{1}{\sqrt{2}}\big(v+h(x)+iG^0(x)\big)
\end{pmatrix},\hspace{0.5cm}
S(x)=v_S+s(x).
\end{align}
The minimization (tadpole) conditions of the scalar potential can be written as
\begin{align}
\mu^2_H&=\lambda_Hv^2+\mu_{HS}v_S+\frac{\lambda_{HS}}{2}v^2_S,\nonumber\\
m^2_S&=-\frac{\mu^3_S}{v_S}-\mu^{\prime}_Sv_S-\lambda_Sv^2_S-\frac{\mu_{HS}}{2}\frac{v^2}{v_S}-\frac{\lambda_{HS}}{2}v^2 \label{treetadpole}.
\end{align}
\begin{figure}[t]
\begin{center}
\begin{tabular}{cc}
\includegraphics[width=6cm]{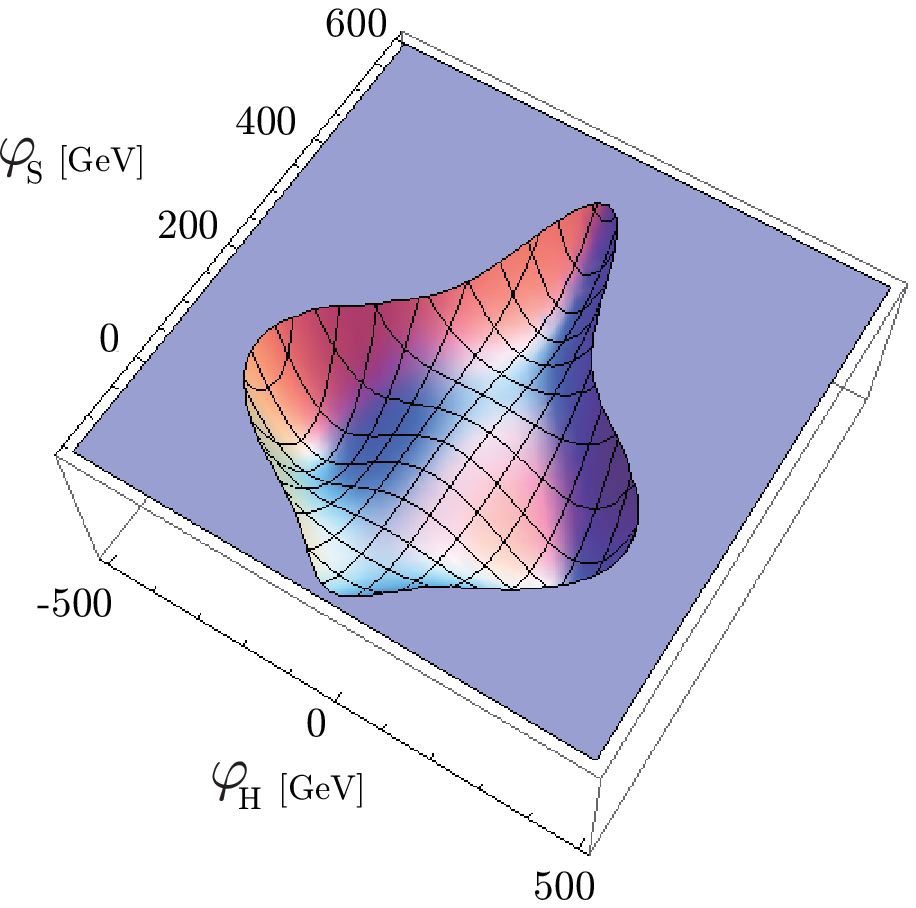}
\hspace{15mm}
\includegraphics[width=6cm]{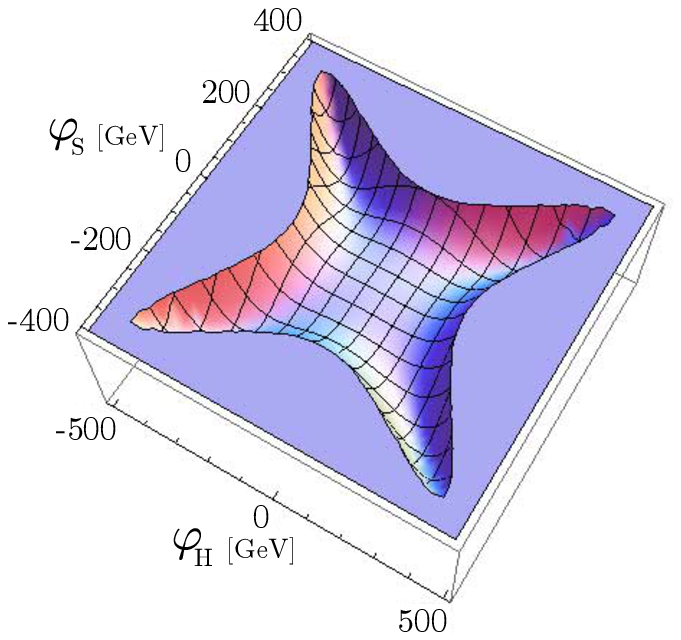}
\end{tabular}
\end{center}
\caption{The shape of the tree-level potential $V_0(\varphi_H,\varphi_S)$ and $V^{Z_2}_0(\varphi_H,\varphi_S)$ as functions of $\varphi_H$ and $\varphi_S$ in left and right figures, respectively. In the left (right) figure, we take $m_{H_1}=125.5 \ {\rm GeV} ,m_{H_2}=150  
\ (500) \ {\rm GeV}, v_S=100 \ (200) \ {\rm GeV}, \alpha=0^{\circ} \ (38^{\circ}),  \mu_{S}^{\prime}=-30 \ (0) \ {\rm GeV}$, and $\mu_{HS}=-80 \ (0) \ {\rm GeV}$.}
\label{tree}
\end{figure}
The mass matrix of $h$ and $s$ (denoted as ${\cal M}^2_H$) has a two-by-two form.
The mass eigenvalues ($m^2_{H_1,H_2}$) are obtained by diagonalizing ${\cal M}^2_H$ 
with an orthogonal matrix $O(\alpha)$,
\begin{align}
{\cal M}^2_H=
\begin{pmatrix}
({\cal M}_H)^2_{11} & ({\cal M}_H)^2_{12}\\
({\cal M}_H)^2_{21} & ({\cal M}_H)^2_{22}
\end{pmatrix}
=
\begin{pmatrix}
\cos\alpha & -\sin\alpha\\
\sin\alpha & \cos\alpha
\end{pmatrix}
\begin{pmatrix}
m^2_{H_1} & 0 \\
0 & m^2_{H_2}
\end{pmatrix}
\begin{pmatrix}
\cos\alpha & \sin\alpha\\
-\sin\alpha & \cos\alpha
\end{pmatrix},
\end{align}
where the value of $\alpha$ is defined in the range $-{\pi}/{4} \leq \alpha \leq {\pi}/{4}$,
and each mass matrix element is given, respectively, by 
\begin{align}
({\cal M}_H)^2_{11}&=2\lambda_Hv^2,\nonumber\\
({\cal M}_H)^2_{12}&=\mu_{HS}v+\lambda_{HS}vv_S,\nonumber\\
({\cal M}_H)^2_{22}&=-\frac{\mu^3_S}{v_S}+\mu^{\prime}_Sv_S+2\lambda_Sv^2_S-\frac{\mu_{HS}}{2}\frac{v^2}{v_S},
\end{align}
where we used the tadpole conditions (\ref{treetadpole}).

The tree-level effective potential takes the form
\begin{align}
V_0(\varphi_H,\varphi_S)=&\frac{\lambda_H}{4}(\varphi_H^4-2v^2_H\varphi^2_H)+\frac{\mu_{HS}}{2}\left(\varphi^2_H\varphi_S-\varphi^2_Hv_S-\frac{v^2_H\varphi^2_S}{2v_S}\right)\nonumber\\
&+\frac{\lambda_{HS}}{4}(\varphi^2_H\varphi^2_S-\varphi^2_Hv^2_S-v^2_H\varphi^2_S)+\mu^3_S\left(\varphi_S-\frac{\varphi^2_S}{2v_S}\right)\nonumber\\
&+\frac{\mu^{\prime}_S}{3}\left(\varphi^3_S-\frac{3}{2}v_S\varphi^2_S\right)+\frac{\lambda_S}{4}(\varphi^4_S-2v^2_S\varphi^2_S),
\end{align}
where $\varphi$ and $\varphi_S$ are the constant background fields of the doublet and singlet Higgsses, respectively. 
The Higgs potential has to satisfy the following conditions to be bounded from below: 
\begin{align}
\lambda_H>0, \hspace{0.5cm} \lambda_{S}>0, 
\hspace{0.5cm} 4\lambda_{H}\lambda_S>\lambda^2_{HS},
\end{align}
where the last condition is needed if $\lambda_{HS}< 0$. If we impose a $Z_2$ symmetry, 
the Higgs potential (\ref{treepotential}) is reduced to
\begin{align}
V^{Z_2}_0=&-\mu^2_HH^{\dagger}H+\lambda_H(H^{\dagger}H)^2+\frac{\lambda_{HS}}{2}H^{\dagger}HS^2+\frac{m^2_S}{2}S^2+\frac{\lambda_S}{4}S^4.
\end{align} 
In Fig.~\ref{tree}, we show representative examples of $V_0(\varphi_H,\varphi_S)$ and $V_0^{Z_2}(\varphi_H,\varphi_S)$. In the left (right) figure, we set $m_{H_1}=125.5 \ {\rm GeV} ,m_{H_2}=150  
\ (500) \ {\rm GeV}, v_S=100 \ (200) \ {\rm GeV}, \alpha=0^{\circ} \ (38^{\circ}),  \mu_{S}^{\prime}=-30 \ (0) \ {\rm GeV}$, and $\mu_{HS}=-80 \ (0) \ {\rm GeV}$. Both Higgs potentials are symmetric about the $\varphi_H$ axis.
On top of this, $V_0^{Z_2}(\varphi_H,\varphi_S)$ is also symmetric about the $\varphi_S$ axis 
because of the $Z_2$ symmetry.
As discussed in Ref.~\cite{Funakubo:2005pu}, the vacuum structures at zero temperature 
may provide some information about the patterns of the phase transitions.
In the left Higgs potential, there is a local minimum on the $\varphi_S$ axis.
As we discuss later, the phase transition can occur twice, i.e., the transition 
from the origin to the local minimum along the $\varphi_S$ axis, 
followed by the transition from there to our vacuum as the temperature decreases.
In the right figure, on the other hand, the phase transition may proceed once, i.e., 
the transition directly from the origin to our vacuum. 
We discuss various patterns of the phase transitions in the next section.

At the tree level, the interactions of $H_1$ and $H_2$ with $Z$ and $W$ bosons are 
\begin{align}
{\cal L}_{\rm HVV}=\frac{1}{v}\left(\cos\alpha \ H_1-\sin\alpha \ H_2\right)\left(2m^2_WW_{\mu}^+W^{-\mu}+m^2_ZZ_{\mu}Z^{\mu}\right),
\end{align}
and the interactions with quarks $f$ are
\begin{align}
{\cal L}_{\rm Yukawa}=-\sum_f\frac{m_f}{v}\left(\cos\alpha \ H_1 -\sin\alpha \ H_2\right)\bar{f}f.
\end{align}
We define the Higgs couplings to gauge bosons and fermions normalized to the corresponding SM ones as
\begin{align}
\kappa_V=\frac{g_{H_1VV}}{g^{\rm SM}_{hVV}}=\cos\alpha,\hspace{1cm}
\kappa_F=\frac{g_{H_1ff}}{g^{\rm SM}_{hff}}=\cos\alpha.
\end{align}
Since $\kappa_V$ and $\kappa_F$ have the same values as in the rSM,
we collectively denote them as $\kappa$ in the following.

\section{Electroweak phase transition}\label{sec:EWPT}
In addition to the tree-level potential (\ref{treepotential}), we include the one-loop Coleman-Weinberg potential at zero temperature~\cite{Coleman:1973jx,Jackiw:1974cv}, which is given by
\begin{align}
V_1(\varphi_H,\varphi_S)=\sum_i n_i \ \frac{\bar{m}^4_i(\varphi_H,\varphi_S)}{64\pi^2}\left(\ln\frac{\bar{m}^2_i(\varphi_H,\varphi_S)}{\mu^2} -c_i \right),
\end{align}
where $\mu$ is a renormalization scale, which will be set at $v$. $\bar{m}_i$ is the background field-dependent mass, and the numerical constant $c_i$ is $3/2$ ($5/6$) for scalars and fermions (gauge bosons). $n_i$ 
are the degrees of freedom of the particle species $i$ $(=H_{1,2},G^0,G^{\pm},W,Z,t,b$), 
which are, respectively, given by
\begin{align}
n_{H_1}=n_{H_2}=n_{G^0}=1, \hspace{0.5cm}
n_{G^{\pm}}=2,\hspace{0.5cm}
n_W=2\cdot 3,\hspace{0.5cm}
n_Z=3,\hspace{0.5cm}
n_t=n_b=-4N_c,\label{degree}
\end{align}
where $N_c$ is the number of colors.

The finite-temperature component of the one-loop effective potential can be written as
\begin{align}
V_1(\varphi_H,\varphi_S,T)=\sum_in_i \ \frac{T^4}{2\pi^2}I_{B,F}\left(\frac{\bar{m}^2_i(\varphi_H,\varphi_S)}{T^2}\right),
\end{align}
where
\begin{align}
I_{B,F}(a^2)=\int^{\infty}_0 dx \ x^2\ln \left(1\mp e^{-\sqrt{x^2+a^2}} \right),
\end{align}
with the upper (lower) sign for bosons (fermions). If $T$ is high compared to $m_i(\varphi_H,\varphi_S)$, $I_{B,F}$ can be expressed as \cite{Dolan:1973qd}
\begin{align}
I_B(a^2)&=-\frac{\pi^4}{45}+\frac{\pi^2}{12}a^2-\frac{\pi}{6}(a^2)^{3/2}-\frac{a^4}{32}\ln\left(\frac{a^2}{\alpha_B}\right)+\cdots, \nonumber\\
I_F(a^2)&=\frac{7\pi^2}{360}-\frac{\pi^2}{24}a^2-\frac{a^4}{32}\ln\left(\frac{a^2}{\alpha_F}\right)+\cdots.\label{highT}
\end{align}
where $\log\alpha_B=2\log4\pi+3/2-2\gamma_E$ and  $\log\alpha_F=2\log\pi+3/2-2\gamma_E$.
Moreover, in order to improve the calculation of the effective potential, 
we include the so-called  daisy contributions \cite{Carrington:1991hz}
\begin{align}
V_{\rm daisy}(\varphi_H,\varphi_S,T)=-\sum_j n_j  \frac{T}{12\pi}\left[\left\{\bar{M}^2_j(\varphi_H,\varphi_S,T)\right\}^{3/2}-\left\{\bar{m}^2_j(\varphi_H,\varphi_S)\right\}^{3/2} \right],
\end{align}
where $\bar{M}^2_j$ are the thermally corrected boson masses
\begin{align}
\bar{M}^2_j(\varphi_H,\varphi_S,T)=\bar{m}^2_j(\varphi_H,\varphi_S)+\Pi_j(T),
\end{align}
and $\Pi_j(T)$ are the finite-temperature mass functions given in Refs.~\cite{Carrington:1991hz,Espinosa:2011ax}. Although the mass-squared values of the scalar and Nambu-Goldstone bosons can be negative, the above daisy contributions can compensate for them at high temperature. 
The full effective potential at finite temperature is given by
\begin{align}
V_{\rm eff}(\varphi_H,\varphi_S,T)=V_0(\varphi_H,\varphi_S)+V_1(\varphi_H,\varphi_S)+V_1(\varphi_H,\varphi_S,T)+V_{\rm daisy}(\varphi_H,\varphi_S,T).
\label{Veff}
\end{align}
For the first-order EWPT, the effective potential (\ref{Veff}) needs to have 
two degenerate minima at the critical temperature $T_C$. The VEVs at $T_C$ are denoted as
\begin{align}
v_C &=\lim_{T\uparrow T_C} v(T_C),\quad
v_{SC}=\lim_{T\uparrow T_C}v_S(T_C),\quad
v_{SC}^{\rm sym}=\lim_{T\downarrow T_C}v_S(T_C),
\end{align}
where the up (down) arrow indicates that $T$ approaches $T_C$ from below (above).
Since the numerical evaluations of $I_B$ and $I_F$ are time consuming, we replace them with the fitting functions adopted in Ref. \cite{Funakubo:2009eg}. The errors of the fitting functions are less than $10^{-6}$ for any $a^2$, which is sufficient for our numerical evaluations.

\begin{figure}[t]
\center
\includegraphics[width=6cm]{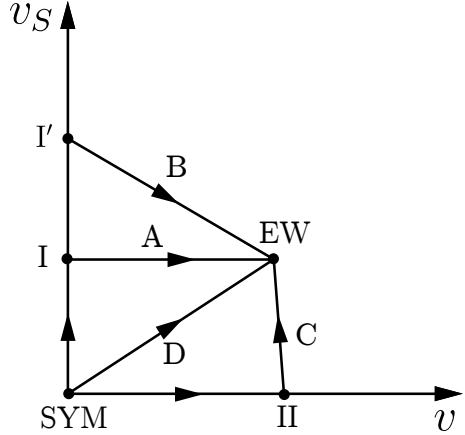} 
\caption{The diverse patterns of the EWPT.}
\label{fig:EWPTpatterns}
\end{figure}

Here, we explain how the EWPT proceeds in the rSM. 
At high temperatures, SU(2)$\times$U(1)$_{Y}$ gauge symmetry is restored. 
However, as the temperature goes down, the symmetric phase (SYM) is no longer the vacuum, 
and eventually the electroweak phase (EW) becomes the global minimum. 
In general, this transition occurs thorugh multiple steps.\footnote{For earlier studies on the multistep EWPT, see, i.e., Ref.~\cite{2stepPT}.}
The various phases that occurs at the intermediate stage are listed in Table \ref{variousphases}. 
As discussed in Ref. \cite{Funakubo:2005pu}, there are four types of transitions. 
We show each path of the transitions in Fig.~\ref{fig:EWPTpatterns}. 
In type C, $v$ initially has a nonzero value, and then $v_S$ starts to increase.  
In type A, the transition occurs from SYM to I phase, followed by the transition I $\to$ EW 
with an almost constant $v_S$. 
In type B, the transition is the same as in type A but $v_S$ varies in the second transition. 
In type D, SYM goes to the EW phase directly.
In principle, we can consider a case in which II phase corresponds to our vacuum. 
However, we do not pursue this case here for simplicity. 

\begin{table}[t]
\begin{center}
\begin{tabular}{|c|c|}
\hline
Phases &  Order parameters \\
\hline\hline
EW & $v= 246$ GeV, $v_S\neq0$  \\
SYM & $v= 0$, $v_S = 0$  \\
\hline
I, I$^{\prime}$ & $v=0, v_S\neq 0$\\
II & $v\neq 0, v_S= 0$\\
\hline
\end{tabular}
\caption{Various phases in the rSM. }
\label{variousphases}
\end{center}
\end{table}

\section{Sphaleron decoupling condition}\label{sec:sphaleron}
In the EWBG mechanism, the BAU is generated through the sphaleron process in the symmetric phase
during the EWPT.
In order to leave the BAU, the sphaleron process has to be decoupled right after the EWPT, namely, the $B$-changing rate in the broken phase ($\Gamma_B^{(b)}$)
must be less than the Hubble constant [$H(T)$]
\begin{align}
\Gamma_B^{(b)}(T)\simeq ({\rm prefactor})e^{-E_{\rm sph}(T)/T} 
< H(T)\simeq 1.66\sqrt{g_*(T)}T^2/m_{\rm P} \label{GamvsH}
\end{align}
where $E_{\rm sph}(T)$ is the sphaleron energy, 
$g_*$ counts the degrees of freedom of relativistic particles in the thermal plasma
($g_*=107.75$ in the rSM), and $m_{\rm P}$ denotes the Planck mass ($1.22\times 10^{19}$ GeV).
The prefactor denotes a fluctuation determinant around the sphaleron (see, e.g., Refs.~\cite{Arnold:1987mh,Hellmund:1991ub,Funakubo:2009eg,D'Onofrio:2012jk}).
From Eq.~(\ref{GamvsH}), it follows that
\begin{align}
\frac{v(T)}{T} > \frac{g_2}{4\pi \mathcal{E}(T)}
\bigg[
	42.97+\ln\mathcal{N} -2\ln\left(\frac{T}{100~{\rm GeV}}\right)+\cdots
\bigg]\equiv \zeta_{\rm sph}(T),\label{sph_dec}
\end{align}
where we use $E_{\rm sph}(T)=4\pi v(T)\mathcal{E}(T)/g_2$,
with $g_2$ being the SU(2) gauge coupling constant. 
$\mathcal{N}$ denotes the translational and rotational zero-mode factors of the fluctuations about the sphaleron.
Note that the dominant contribution in $\zeta_{\rm sph}(T)$ comes from $\mathcal{E}(T)$
while the next-to-leading term may be $\ln\mathcal{N}$.
In the MSSM, $\ln\mathcal{N}$ typically amounts to 10\%~\cite{Funakubo:2009eg}.
The last term can become relevant if $T_C$ is significantly lower than 100 GeV.
As is well known, this can happen in the rSM.
In the current investigation, we exclusively focus on the calculation of $\mathcal{E}(T)$ by using Eq.({\ref {Veff}}) and evaluate Eq.({\ref {sph_dec}}) without the $\ln\mathcal{N}$ term as a first step toward the complete analysis.
It should be noticed that the EWPT starts to develop after the nucleation of Higgs bubbles. It is thus better to evaluate Eq.(25) at such a nucleation temperature ($T_N$) rather than at $T_C$. 
In this paper, however, we adopt $v(T_C)/T_C>\zeta_{\rm sph}(T_C)$ as the sphaleron decoupling criterion for simplicity.

We closely follow the method given in Refs.~\cite{Manton:1983nd,Klinkhamer:1984di,Funakubo:2005bu}
to obtain the sphaleron solution (for earlier studies on the sphaleron solutions in the rSM, see, i.e., Ref.~\cite{sph_rSM}).
Since ${\rm U(1)}_Y$ contributions are sufficiently small~\cite{sph_wU1Y}, we employ
the spherically symmetric ansatz. 
Specifically, we consider the configuration space spanned by the following: 
\begin{align}
 A_i(\mu,r,\theta,\phi)
 &=-\frac{i}{g_2}f(r)\partial_iU(\mu,\theta,\phi)U^{-1}(\mu,\theta,\phi),\\ 
H(\mu,r,\theta,\phi)
&=\frac{v}{\sqrt{2}}\left[(1-h(r))
	\left(
	\begin{array}{c}
	0 \\
	e^{-i\mu}\cos\mu
	\end{array}
	\right)+h(r)U(\mu,\theta,\phi)
	\left(
	\begin{array}{c}
	0 \\
	1
	\end{array}
	\right)\right], \\
S(\mu, r, \theta, \phi) &=v_Sk(r),
\end{align}
where $A_i$ are SU(2) gauge fields, and $U$ is defined as
\begin{eqnarray}
U(\mu,\theta,\phi)=
	\left(
	\begin{array}{cc}
	e^{i\mu}(\cos\mu-i\sin\mu\cos\theta) & e^{i\phi}\sin\mu\sin\theta \\
	-e^{-i\phi}\sin\mu\sin\theta & e^{-i\mu}(\cos\mu+i\sin\mu\cos\theta)
	\end{array}
	\right),
\end{eqnarray}
with $\mu \in [0, \pi]$. 
$U(\mu, \theta, \phi)$ is noncontractible since $\pi_3({\rm SU}(2))\simeq \mathbb{Z}$.
$\mu$ parametrizes the least-energy path connecting the topologically 
different vacua. The configuration at $\mu=\pi/2$ corresponds to the sphaleron.

The energy functional in $A_0=0$ is given by
\begin{eqnarray}
E[H, S]=\int d^3\boldsymbol{x}
\left[
	\frac{1}{4}F^a_{ij}F^a_{ij}+(D_iH)^\dagger D_iH
	+\frac{1}{2}\partial_i S\partial_i S+V_{\rm eff}(H, S, T)
\right],
\end{eqnarray}
where the tree-level potential is replaced with the one-loop corrected one in order to incorporate the high-order effects. 
For $\mu=\pi/2$, one gets
\begin{align}
E_{\rm sph}[f,h,k] 
&= \frac{4\pi v}{g_2}\int^\infty_0 d\xi~
\bigg[
	4\left(\frac{df}{d\xi}\right)^2+\frac{8}{\xi^2}(f-f^2)^2
	+\frac{\xi^2}{2}\left(\frac{dh}{d\xi}\right)^2
	+h^2(1-f)^2 \non\\
&\hspace{7cm}+\frac{\xi^2}{2}\frac{v_S^2}{v^2}\left(\frac{dk}{d\xi}\right)^2
	+\frac{\xi^2}{g_2^2v^4}V_{\rm eff}(h, k, T)
\bigg],\label{Esph}
\end{align}
where $\xi=g_2 v r$. From Eq.~(\ref{Esph}), the equations of motion are found to be
\begin{align}
\frac{d^2f}{d\xi^2} 
&= \frac{2}{\xi^2}f(1-f)(1-2f)-\frac{1}{4}h^2(1-f),\\
\frac{d}{d\xi}\left(\xi^2\frac{dh}{d\xi}\right)
&= 2h(1-f)^2+\frac{\xi^2}{g_2^2}\frac{1}{v^4}\frac{\partial V_{\rm eff}}{\partial h}, \\
\frac{d}{d\xi}\left(\xi^2\frac{dk}{d\xi}\right)
&= \frac{\xi^2}{g_2^2}\frac{1}{v^2v_S^2}\frac{\partial V_{\rm eff}}{\partial k}.
\end{align}
We solve the above equations with the following boundary conditions
\begin{eqnarray}
&&\lim_{\xi\to0} f(\xi) = 0,\quad \lim_{\xi\to0} h(\xi) = 0,\quad \lim_{\xi\to0} k'(\xi) = 0,  \\
&&\lim_{\xi\to\infty} f(\xi) = 1,\quad \lim_{\xi\to\infty} h(\xi) = 1,\quad \lim_{\xi\to\infty} k(\xi) = 1.
\end{eqnarray}

Before going into the detailed analysis of the EWPT and the sphaleron decoupling condition,
we demonstrate how $\mathcal{E}$ depends on model parameters.
At this point, we use the zero-temperature one-loop effective potential to extract the non-temperature
dependence, and the full analysis is performed in Sec.~\ref{sec:numerics}.

\begin{figure}[t]
\center
\includegraphics[width=8cm]{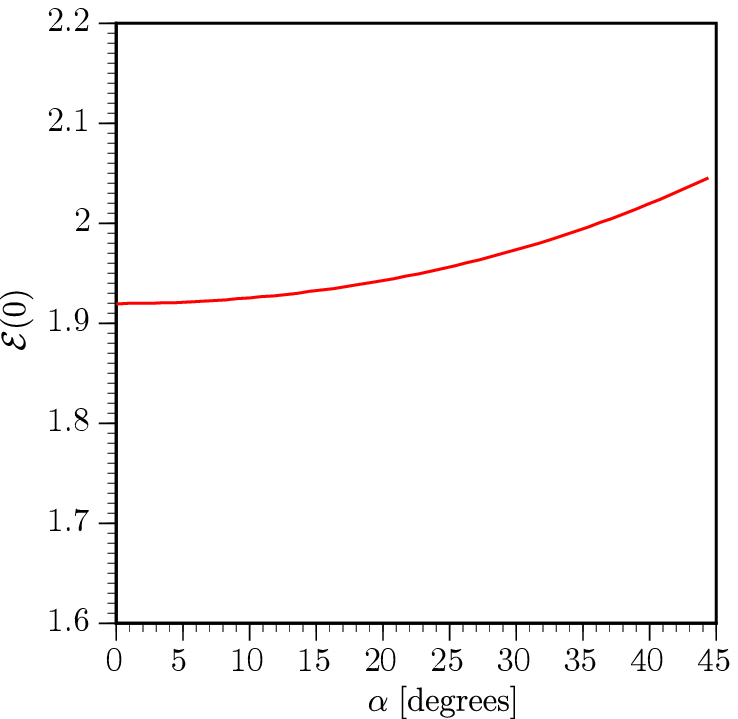} 
\hspace{0.5cm}
\includegraphics[width=7.5cm]{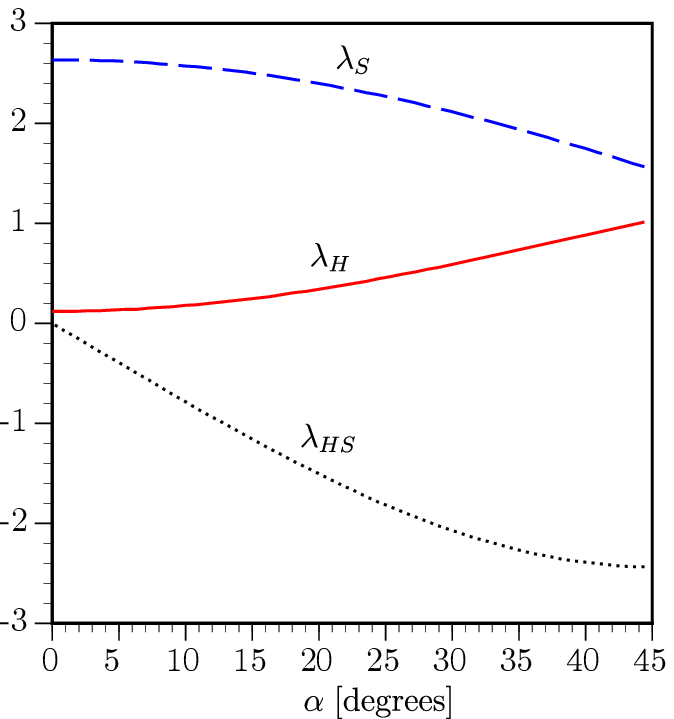}
\caption{The sphaleron energy (left panel) and $\lambda_{H,S,HS}$ (right panel) are plotted as a function of $\alpha$. We set $m_{H_1}=125.5$ GeV, $m_{H_2}=500$ GeV, $v_S=200$ GeV and $\mu_S=\mu^{\prime}_{S}=\mu_{HS}=$ 0 GeV.}
\label{fig:Esph_ex}
\end{figure}

In the left panel of Fig.~\ref{fig:Esph_ex}, $\mathcal{E}$ is shown as a function of $\alpha$. 
We take $m_{H_1}=125.5$ GeV, $m_{H_2}=500$ GeV, $v_S=200$ GeV and $\mu_S=\mu^{\prime}_{S}=\mu_{HS}=$ 0 GeV as an example.
We can see that $\mathcal{E}$ gets larger as $\alpha$ increases.  
It rises about 5\% from $\alpha=0^\circ$ to $\alpha\simeq 45^\circ$.
To understand this behavior, we also show $\lambda_H$, $\lambda_S$ and $\lambda_{HS}$
in the right panel of Fig.~\ref{fig:Esph_ex}.
These couplings are determined after fixing $m_{H_{1,2}}$ and $\alpha$.
$\lambda_H$ has to increase according to the rise of $\alpha$ to keep the value of $m_{H_1}$ fixed. 
On the other hand, $\lambda_S$ behaves oppositely, as it should.
We can see that the increment of $\mathcal{E}$ is due to the enhancement of $\lambda_H$. 
This correlation has already been shown within the SM~\cite{Klinkhamer:1984di}, 
and the same behavior is observed in the rSM.
We also find a mild dependence of $m_{H_2}$ on $\mathcal{E}(0)$, 
which is again essentially because of the increase or decrease of $\lambda_H$. 

\section{Numerical analysis}\label{sec:numerics}
In this section, we perform the numerical analysis.
In the rSM, there are eight parameters in the tree-level potential:
\begin{align}
\mu_H^2,~m_S^2,~\mu_S,~\mu'_S,~\mu_{HS},~\lambda_H,~\lambda_{HS},~\lambda_S.
\end{align}
In our analysis, $\mu_{H}^2$, $m_S^2$, $\lambda_H$, $\lambda_S$ and $\lambda_{HS}$
are replaced by $v$, $v_S$, $m_{H_1}$, $m_{H_2}$ and $\alpha$. 
This replacement can be done by solving the following coupled equations:
\begin{align}
\frac{1}{v}\left\langle\frac{\partial V_{\rm eff}}{\partial \varphi_H}\right\rangle =
\frac{1}{v_S}\left\langle\frac{\partial V_{\rm eff}}{\partial \varphi_S}\right\rangle &=0, 
\label{tad_1L}\\
(\mathcal{M}_H^2)_{11}-m_{H_1}^2\cos^2\alpha-m_{H_2}^2\sin^2\alpha &=0,\label{m11_1L}\\
(\mathcal{M}_H^2)_{22}-m_{H_1}^2\sin^2\alpha-m_{H_2}^2\cos^2\alpha &=0,\label{m22_1L}\\
(\mathcal{M}_H^2)_{12}-(m_{H_1}^2-m_{H_2}^2)\sin\alpha\cos\alpha &=0\label{m12_1L}.
\end{align} 
where we take $m_{H_1}=125.5$ GeV and $v = 1/(2^{1/4}\sqrt{G_F})(\simeq 246$ GeV) with
$G_F$ being the Fermi coupling constant.
$\langle X\rangle$ denotes that $X$ is evaluated in the vacuum, $\varphi_H=v$ and $\varphi_S=v_S$. 
Since the Nambu-Goldstone boson loop contributions that have to be treated with some care
are numerically unimportant, we will not take them into account in the current investigation.
Throughout our analysis, we take $\mu_S=0$. 

As mentioned in the Introduction, the EWPT in the SM is not strongly first order for the observed Higgs mass. 
In order to circumvent this, the Higgs potential has to be altered 
by the doublet-singlet Higgs mixing terms (H-S mixing parameters : $\mu_{HS}$ and $\lambda_{HS}$).  
Here, we consider the following two cases:
\begin{itemize}
\item[(i)] One H-S mixing parameter [($\lambda_{HS}\neq0$, $\mu_{HS}=0$) 
or ($\lambda_{HS}=0$, $\mu_{HS}\neq0$)],
\item[(ii)] Two H-S mixing parameters [$\lambda_{HS}\neq 0$, $\mu_{HS}\neq0$].
\end{itemize}
Without going into the detailed investigation, we can foresee that the latter case has a larger window to the strong first-order EWPT. We conduct the analysis based on the improved sphaleron decoupling condition.

%
%
\subsection{One H-S mixing parameter case}
\begin{figure}[t]
\center
\includegraphics[width=8cm]{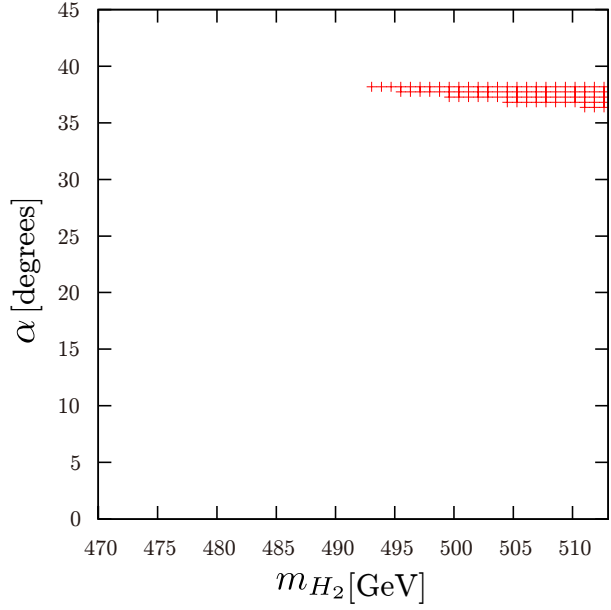}
\caption{The possible region for the strong first-order phase transition in Case (i). In this figure, we set $m_{H_1}=125.5 \ {\rm GeV}$, $v_S=200 \ {\rm GeV}$, and $\mu_S^{\prime}=\mu_{HS}=\mu_S = 0$ GeV.}
\label{distribution_Ia}
\end{figure}

Here, we consider Case (i) with $\lambda_{HS}\neq0$ and $\mu_{HS}=0$.\footnote{A representative example with $\lambda_{HS}=0$ and $\mu_{HS}\neq0$  is given in Table~\ref{tab:bm_summary}.}
In Fig. \ref{distribution_Ia}, we show an allowed region where the following condition is achieved in the  ($m_{H_2}$,$\alpha$) plane:
\begin{align}
\frac{v_C}{T_C} > \zeta_{\rm sph}(T_C).
\end{align}
We here take $m_{H_1}=125.5 \ {\rm GeV}$, $v_S=200 \ {\rm GeV}$, and $\mu_S^{\prime}=\mu_{HS}= 0$ GeV.\footnote{We allow a small $Z_2$-breaking term to avoid the domain wall problem~\cite{Zeldovich:1974uw},
so what we call the $Z_2$ model is an approximate one.}
It is found that only large values of $\alpha$ and $m_{H_2}$  are allowed. In addition, the sign of $\alpha$ needs to be positive --- and thus $\lambda_{HS}$ is negative --- for the first-order phase transition as discussed in Ref. \cite{Profumo:2007wc}. However, these large values of $\alpha$ and $m_{H_2}$ receive stringent constraints 
from the EW precision tests \cite{Peskin:1990zt,Maksymyk:1993zm}. According to Ref. \cite{Baek:2012uj}, $\alpha$ is less than about 23$^{\circ}$ for $m_{H_2} \gtrsim 400 $  GeV.
Moreover, the recent LHC data indicates that the value of $\kappa$ is bounded as follows \cite{ATLAS,CMS}: 
\begin{align}
\kappa_V=1.15^{+0.08}_{-0.08}~({\rm ATLAS}),\hspace{0.5cm}
\kappa_V=0.81-0.97~({\rm CMS}).
\end{align}  
Thus, it seems difficult to have the strong first-order phase transition 
in the spontaneously broken $Z_2$ model.
On the other hand, there may be a viable window in the unbroken $Z_2$ model. If we consider the II phase as the EW phase, $\alpha=0$ in such a vacuum since $v_S=0$. In this case, the transition SYM$\to$ I(I$'$)$\to$ II can induce the strong first-order phase transition, so the above experimental constraints 
are no longer stringent. 
For a recent study on such a possibility, see, for instance, Ref. \cite{Cline:2012hg}.

\begin{figure}[t]
\center
\includegraphics[width=7.5cm]{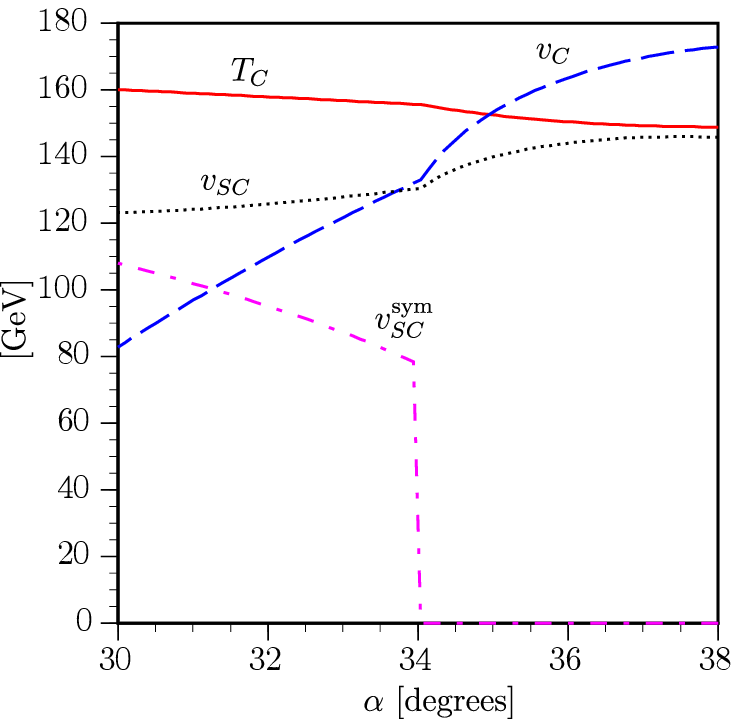} \hspace{0.5cm}
\includegraphics[width=7cm]{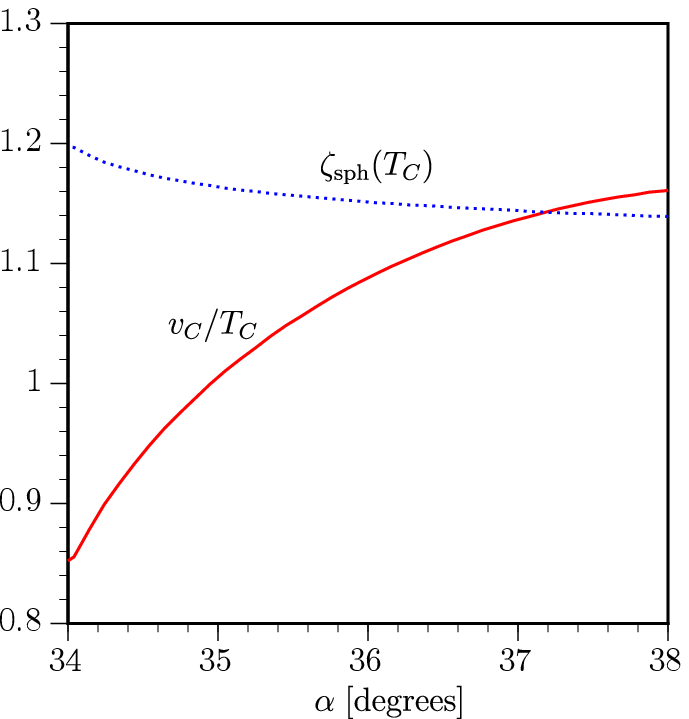}
\caption{The Higgs VEVs at $T_C$ (left panel), and $v_C/T_C$ and $\zeta_{\rm sph}$ (right panel) are presented as functions of $\alpha$ in Case (i).}
\label{fig:1stPT_Ia}
\end{figure}

Although the successful region in the $Z_2$ model is strongly disfavored
by the LHC data and EW precision tests, we give an example that shows how the EWPT is strengthened
and which type of the EWPT can be realized. More importantly, 
the typical size of $\zeta_{\rm sph}(T_C)$ in the rSM is epitomized by this example.
An example is shown in Fig.~\ref{fig:1stPT_Ia}, in which $m_{H_2}=500$ GeV is chosen.
In the left panel, the Higgs VEVs at $T_C$ are displayed. 
At around $\alpha=34^\circ$, $v_{SC}^{\rm sym}$ becomes zero, so the type of the EWPT 
 changes from B to D . 
It is found that the conventional decoupling criterion $v_C/T_C>1$ is satisfied 
for $\alpha\gtrsim 35^\circ$.
The right panel shows $v_C/T_C$ and $\zeta_{\rm sph}(T_C)$ in the narrowed $\alpha$ range.
The crossing point occurs at around $\alpha= 37.2^\circ$, which is somewhat stronger than
the bound obtained by the conventional criterion. 
We also find that for $\alpha\gtrsim 38^\circ$, the EWPT is changed into type C which is weakly first order, as expected. 
In summary, the EWPT in Case (i) can be strongly first order 
due to the sizable $\alpha$ (or equivalently, due to the sizable $\lambda_{HS}$ with a negative sign), and its PT property is type D.
%
%
\subsection{Two H-S mixing parameters case}
\begin{figure}[t]
\center
\includegraphics[width=8cm]{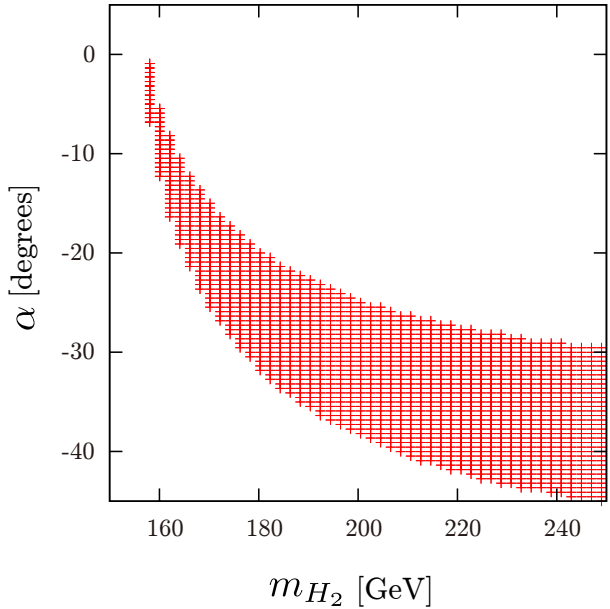}
\caption{The possible region for the strong first-order phase transition in Case (ii). In this figure, we set $m_{H_1}=125.5 \ {\rm GeV}$, $v_S=90 \ {\rm GeV}$, $\mu_S^{\prime}=-30 \ {\rm GeV}$, $\mu_{HS}=-80 \ {\rm GeV}$ and $\mu_S$= 0 \ GeV.}
\label{distribution_II}
\end{figure}

We move on to analyze Case (ii). Figure \ref{distribution_II} shows the possible allowed region for the strong first-order phase transition in Case (ii). In this figure, we set $m_{H_1}=125.5 \ {\rm GeV}$, $v_S=90 \ {\rm GeV}$, $\mu_S^{\prime}=-30 \ {\rm GeV}$, and $\mu_{HS}=-80 \ {\rm GeV}$. As you can see, Fig. 6 has larger possible regions  compared to those in the $Z_2$ model. It is found that there is an upper limit of $m_{H_2}$ if we fix $\alpha$. For example, at $\alpha=-15.6^{\circ}$, the successful range of $m_{H_2}$ is less than about $170 \ {\rm GeV}$. Further, $\alpha$ is restricted to the negative region in order to realize the strong first-order phase transition in the chosen parameter set. 

\begin{figure}[t]
\center
\includegraphics[width=7.5cm]{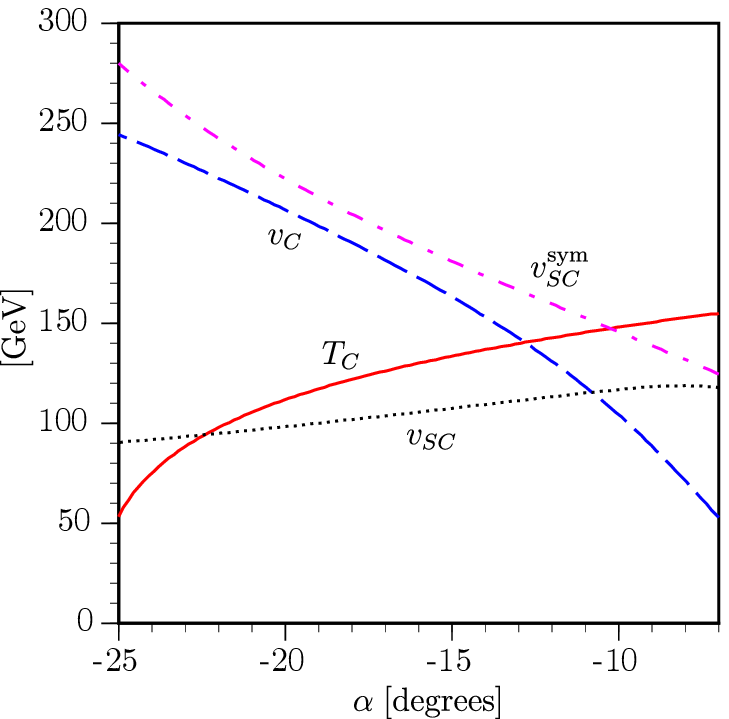} 
\hspace{0.5cm}
\includegraphics[width=7cm]{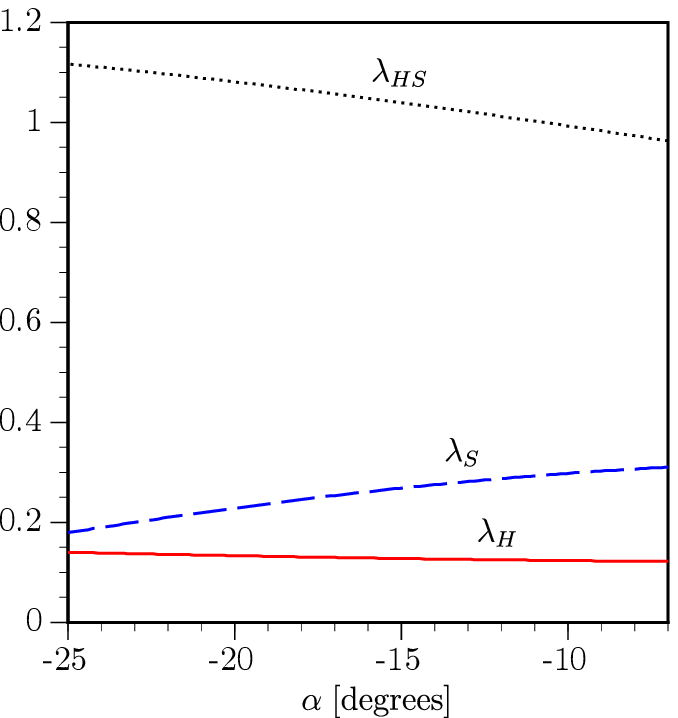} 
\caption{The Higgs VEVs at $T_C$ (left panel) and $\lambda_{H,S,HS}$ (right panel) are shown as functions of $\alpha$ in Case (ii).}
\label{fig:EWPT_II}
\end{figure}

We now look at the EWPT properties in Case (ii) in more detail. 
In the left panel of Fig.~\ref{fig:EWPT_II}, the VEVs at $T_C$ are plotted 
in the range $-25^\circ\leq \alpha\leq -7^\circ$, with $m_{H_2}=170$ GeV.
We can see that $v_C/T_C$ becomes enhanced as $\alpha$ decreases.
In this region, $|v_{SC}^{\rm sym}-v_{SC}|$ is significantly large, so the PT property is type B. Conversely, $|v_{SC}^{\rm sym}-v_{SC}|$ gets smaller for smaller $\alpha$, 
and the PT is eventually reduced to type A which is weakly first order, as explicitly shown here.
In the right panel of Fig.~\ref{fig:EWPT_II}, we plot $\lambda_H$, $\lambda_S$, and $\lambda_{HS}$
in the same $\alpha$ range. A relatively large $\lambda_{HS}$ is needed to drive the EWPT to be
strongly first order. Unlike the $Z_2$ model, the sign of $\lambda_{HS}$ is not necessarily negative
in this case. However, we should note that $\mu_{HS}$ has to be negative. 

\begin{figure}[t]
\center
\includegraphics[width=7.5cm]{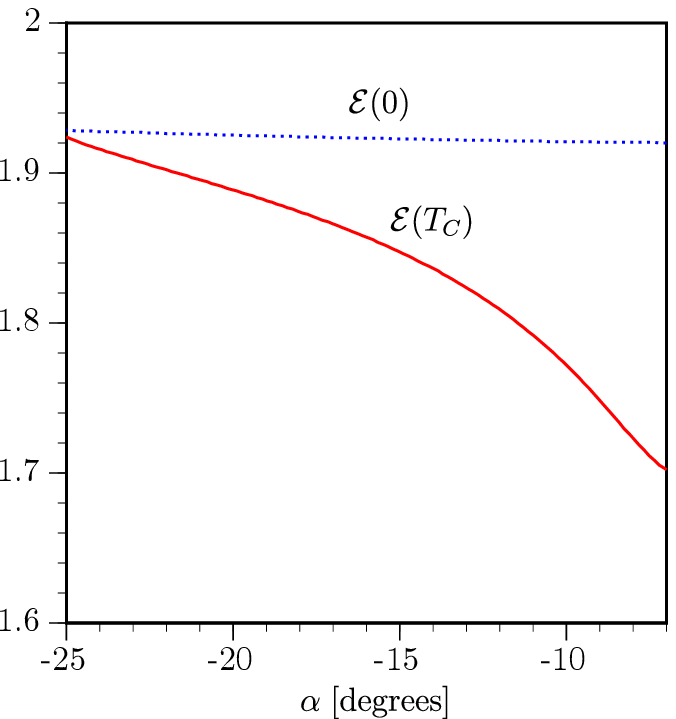} 
\hspace{0.5cm}
\includegraphics[width=7.6cm]{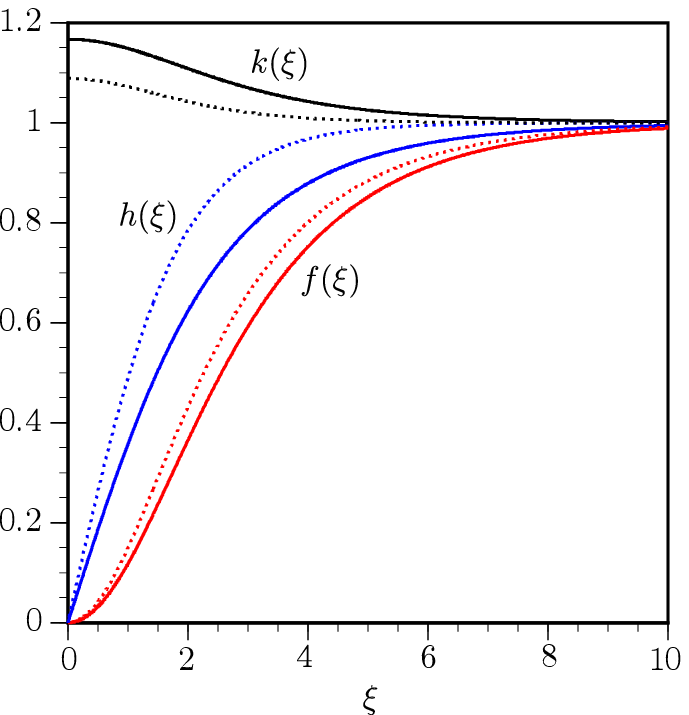}
\caption{(Left panel) $\mathcal{E}(T_C)$ and $\mathcal{E}(0)$ are shown as functions of $\alpha$. (Right panel) The sphaleron profiles are plotted as functions of $\xi$ at $T=0$  GeV and $T=T_C$, where we take $\alpha=-10^\circ$. The straight (dotted) lines  denote $f(\xi)$, $h(\xi)$, and $k(\xi)$ at $T=T_C (0)$ GeV.}
\label{fig:sph_II}
\end{figure}

Now we discuss $\zeta_{\rm sph}$ at the above benchmark point.
We first scrutinize the sphaleron energy with and without a temperature effect.  
In the left panel of Fig.~\ref{fig:sph_II}, we give $\mathcal{E}(0)$ and $\mathcal{E}(T_C)$.
The slight rise of $\mathcal{E}(0)$ with a decreasing $\alpha$ is mainly due to the corresponding
value of $\lambda_H$ which has the same rising behavior as shown in the right panel 
of Fig.~\ref{fig:EWPT_II}.
On the other hand, $\mathcal{E}(T_C)$ has a significant temperature dependence.
It is found that $\mathcal{E}(T_C)<\mathcal{E}(0)$. The simple argument is as follows.
In general, at finite temperatures the energy difference between the broken phase 
and the symmetric phase ($v=0$ GeV) is smaller than that at zero temperature, namely, 
$V_{\rm eff}$ appearing in Eq.~(\ref{Esph}) gets smaller as the temperature increases.
Correspondingly, the gradient energy becomes smaller to balance the potential energy 
in order for the classical solution to exist. 
The right panel of Fig.~\ref{fig:sph_II} quantitatively supports the above argument.
Here, we plot $f(\xi)$, $h(\xi)$, and $k(\xi)$ at $T=0$ GeV and $T_C$, choosing $\alpha=-10^\circ$.
The straight (dotted) lines denote $f(\xi)$, $h(\xi)$, and $k(\xi)$ at $T=T_C(0)$.
As can be seen, the gradients of the doublet Higgs and the gauge field
are lowered at $T_C$. On the other hand, the singlet Higgs does not yield a significant contribution to the gradient energy
in Eq.~(\ref{Esph}).

\begin{figure}[t]
\center
\includegraphics[width=8cm]{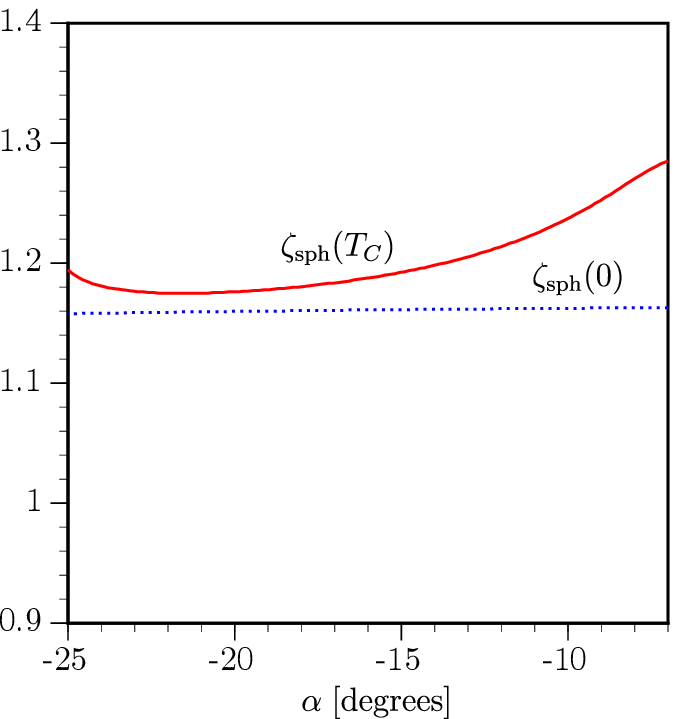}
\hspace{0.5cm}
\includegraphics[width=7.5cm]{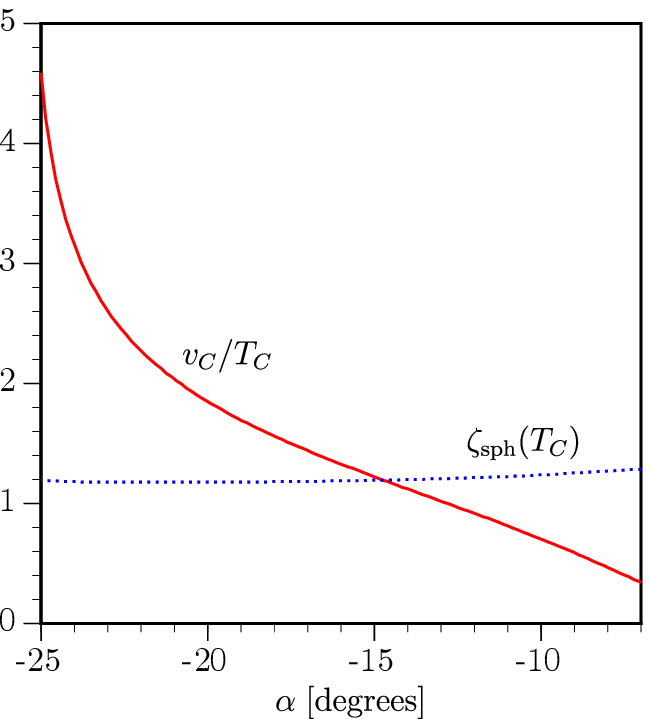}
\caption{(Left panel) $\zeta_{\rm sph}$ at $T_C$ and 0 are plotted as functions of $\alpha$. 
(Right panel) $v_C/T_C$ vs $\zeta_{\rm sph}$ in the same $\alpha$ range as in the left panel.}
\label{fig:1stPT}
\end{figure}

Now let us estimate $\zeta_{\rm sph}(T)$ with the sphaleron energies obtained above.
The left panel of Fig.~\ref{fig:1stPT} shows $\zeta_{\rm sph}(T_C)$ and $\zeta_{\rm sph}(0)$
as a function of $\alpha$.
As discussed in Sec.~\ref{sec:sphaleron}, the smaller $\mathcal{E}$ gives the larger $\zeta_{\rm sph}$.
The slight rise at around $\alpha=-25^\circ$ is due to the third term on the right-hand side 
of Eq.~(\ref{sph_dec}).
In this region, $T_C$ rapidly gets lowered and can be as small as around 50 GeV (as shown 
in the left panel of Fig.~\ref{fig:EWPT_II}), giving rise to some effect.

The $v_C/T_C>\zeta_{\rm sph}(T_C)$ region is presented in the right panel of Fig.~\ref{fig:1stPT}.
$\alpha\lesssim-14.7^\circ$ is needed to achieve the successful sphaleron decoupling.
On the other hand, $\alpha\lesssim-12.9^\circ$ is obtained if $v_C/T_C>1$ is used.


%
%
\subsection{Triple Higgs boson coupling}
We investigate a correlation between the triple Higgs boson coupling ($\lambda_{H_1H_1H_1}$) 
and the strength of the strong first-order EWPT (for a tree-level analysis in the rSM, see Ref. \cite{Noble:2007kk}).
To do so, we evaluate $\lambda_{H_1H_1H_1}$ in the effective potential approach. 
Although the external momentum dependence is inherently important, 
we do not pursue this analysis in this paper and defer it to future work.
$\lambda_{H_1H_1H_1}^{\rm rSM}$ at the one-loop order is given by
\footnote{$\lambda_{H_1H_1H_1}$ in the SM with two real singlets is discussed 
in Ref.~\cite{Ahriche:2013vqa}.}
\begin{align}
\lambda_{H_1H_1H_1}^{\rm rSM} = \lambda_{H_1H_1H_1}^{\rm rSM, tree} 
+\lambda_{H_1H_1H_1}^{\rm rSM, loop}, 
\end{align}
where
\begin{align}
\lambda_{H_1H_1H_1}^{\rm rSM, tree} &=6
\left[
	\lambda_Hvc_\alpha^3
	+\frac{\mu_{HS}}{2}s_\alpha c_\alpha^2
	+\frac{\lambda_{HS}}{2}s_\alpha c_\alpha(vs_\alpha+v_Sc_\alpha)
	+\left(\frac{\mu_S'}{3}+\lambda_Sv_S\right)s_\alpha^3
\right],\\
\lambda_{H_1H_1H_1}^{\rm rSM, loop} &=
c_\alpha^3\left\langle\frac{\partial^3 V_1}{\partial \varphi_H^3}\right\rangle
+c_\alpha^2s_\alpha\left\langle\frac{\partial^3 V_1}{\partial \varphi_H^2\partial \varphi_S}\right\rangle
+c_\alpha s_\alpha^2\left\langle\frac{\partial^3 V_1}{\partial \varphi_H\partial \varphi_S^2}\right\rangle
+s_\alpha^3\left\langle\frac{\partial^3 V_1}{\partial \varphi_S^3}\right\rangle,
\end{align}
where $c_{\alpha}=\cos\alpha$ and $s_{\alpha}=\sin\alpha$.
It should be noted that the vacuum and Higgs boson masses have to be renormalized 
at the one-loop level in order to evaluate $\lambda_{H_1H_1H_1}^{\rm rSM}$ properly;
in our analysis, Eqs.~(\ref{tad_1L}) -- (\ref{m12_1L}) are used for this.
One can easily work out $\lambda_{H_1H_1H_1}$ in the SM 
in the effective potential approach (see, e.g., Ref. \cite{lamhhh}).
In this case, $\lambda_{H_1H_1H_1}$ has the simple form
\begin{align}
\lambda_{H_1H_1H_1}^{\rm SM} 
= \frac{3m_{H_1}^2}{v}
\left[
	1+\frac{9m_{H_1}^2}{32\pi^2v^2}+\sum_{i=W,Z,t,b}n_i\frac{m_i^4}{12\pi^2m_{H_1}^2v^2}
\right]\simeq 175.83~[{\rm GeV}].
\end{align}
Note that the largest one-loop contribution comes from the top-quark loop and it grows as
$\mathcal{O}(m_t^4)$. Such a nondecoupling effect can also appear in the radiative corrections
to the triple Higgs boson coupling in the two Higgs doublet model~\cite{lamhhh}
and some classes of the supersymmetric models~\cite{Kanemura:2012hr}.

We define the deviation of the triple Higgs boson coupling from the SM value as
\begin{align}
\Delta \lambda_{H_1H_1H_1} 
= \frac{\lambda_{H_1H_1H_1}^{\rm rSM}-\lambda_{H_1H_1H_1}^{\rm SM} }{\lambda_{H_1H_1H_1}^{\rm SM}}.
\end{align}

\begin{figure}[t]
\center
\includegraphics[width=8cm]{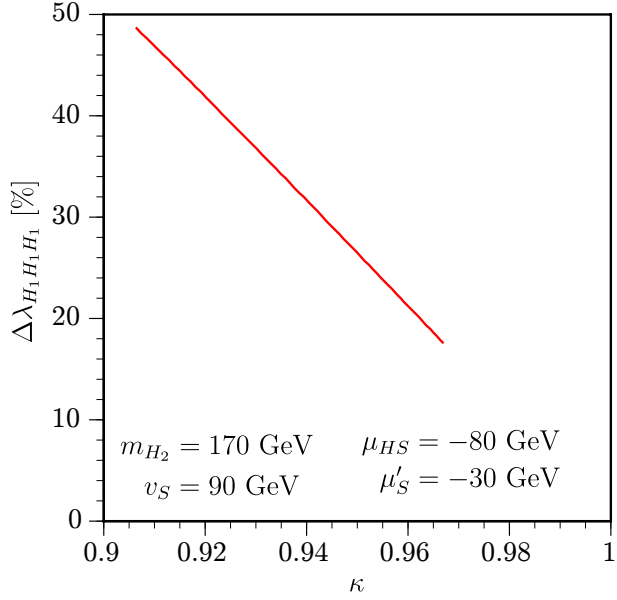}
\caption{$\kappa$ and $\Delta\lambda_{H_1H_1H_1}$ in the strong first-order EWPT region.
The right end point of the red line corresponds to $v_C/T_C\simeq1.20$ and $\zeta_{\rm sph}(T_C)\simeq1.19$,
while $v_C/T_C\simeq4.59$ and $\zeta_{\rm sph}(T_C)\simeq1.19$ at the left end point.}
\label{fig:kappa_lamhhh}
\end{figure}

In Fig.~\ref{fig:kappa_lamhhh}, a relationship between $\kappa$ and 
$\Delta\lambda_{H_1H_1H_1}$ is shown in the strong first-order EWPT region. 
The input parameters are the same as those in Fig.~\ref{fig:EWPT_II}.
It is found that $\kappa\lesssim 0.97$ and $\Delta\lambda_{H_1H_1H_1}\gtrsim17.5~\%$,
where the lower values correspond to the case with $v_C/T_C\simeq 1.20$ and  $\zeta_{\rm sph}(T_C)\simeq 1.19$.
$\Delta\lambda_{H_1H_1H_1}$ can be as large as 48.7\%
at $\kappa\simeq 0.91$, where $v_C/T_C\simeq 4.59$ and $\zeta_{\rm sph}(T_C)\simeq 1.19$, as shown in the right panel of Fig.~\ref{fig:1stPT}.

\begin{figure}[t]
\center
\includegraphics[width=8.7cm]{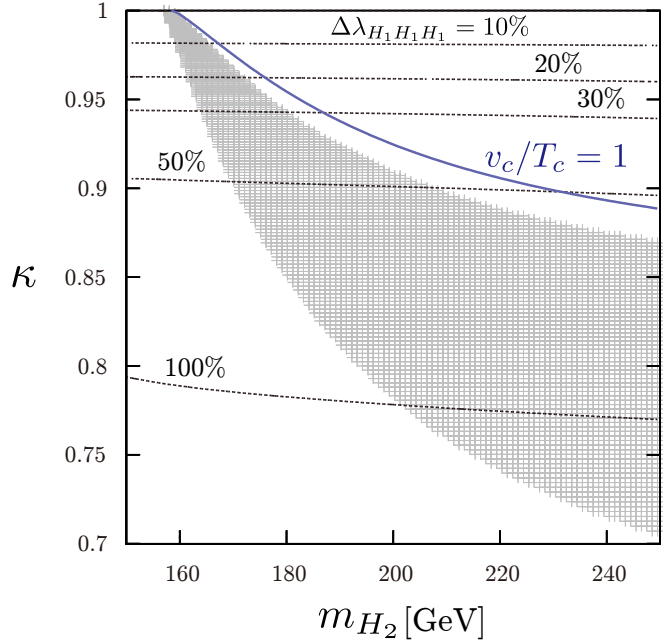}
\caption{$\Delta\lambda_{H_1H_1H_1}$ in the ($m_{H_2}, \kappa$) plane. 
The input parameters are the same as in Fig.~\ref{distribution_II}. 
$v_C/T_C>\zeta_{\rm sph}(T_C)$ is satisfied in the shaded region. 
For a reference, the $v_C/T_C=1$ contour line is also shown.}
\label{fig:comb_II}
\end{figure}

We study $\Delta\lambda_{H_1H_1H_1}$ in larger regions by varying $m_{H_2}$.
Our finding is presented in Fig.~\ref{fig:comb_II}.
$\Delta\lambda_{H_1H_1H_1}$ is overlaid on the same plane as in Fig.~\ref{distribution_II} but with 
$\kappa$ rather than $\alpha$ for the vertical axis.
$\Delta\lambda_{H_1H_1H_1}=$10, 20, 30, 50, and 100\% contours are plotted with the dotted lines from top to bottom.
To emphasize how much the sphaleron decoupling condition is improved, the $v_C/T_C=1$ line is also displayed. For some representative points in the successful EWBG region, we find the following: 
\begin{itemize}
\item $\Delta\lambda_{H_1H_1H_1}\simeq 16\%$ for $\kappa \simeq 0.97$ 
and $160$ GeV $\lesssim m_{H_2}\lesssim$ 169 GeV. 
\item $\Delta\lambda_{H_1H_1H_1}\simeq 27\%$ for $\kappa \simeq 0.95$,
and $163$ GeV $\lesssim m_{H_2}\lesssim$ 176 GeV. 
\item $\Delta\lambda_{H_1H_1H_1}\simeq 50\%$ for $\kappa \simeq 0.90$,
and $170$ GeV $\lesssim m_{H_2}\lesssim$ 206 GeV. 
\end{itemize}
It should be emphasized that the upper bound of $m_{H_2}$ in each case is enlarged to
172, 182, and 227 GeV, respectively, if $v_C/T_C>1$ is used as the sphaleron decoupling condition.
On the other hand, for a fixed $m_{H_2}$ with $\kappa<0.9$, 
$\Delta\lambda_{H_1H_1H_1}$ gets reduced by up to about 8\%
once we adopt $v_C/T_C>1$. 

Here, we comment on the Landau pole issue.
As we have studied in this section, to satisfy the sphaleron decoupling condition 
the H-S mixing parameter ($\lambda_{HS}$) has to be greater than
 certain values, which may hit the Landau pole below the grand unification scale ($\simeq 2\times 10^{16}$ GeV).
To examine this, we solve the renormalization group equations of $\lambda_{H,S, HS}$
(for one-loop $\beta$ functions, see, i.e., Ref. \cite{Gonderinger:2009jp}).
We determine the cutoff ($\Lambda$) of the model by a scale at which $\lambda_{H,S,HS}(\Lambda)>4\pi$ occurs.
In our explored parameter space, $\Lambda$ cannot reach the grand unification scale  
in the successful EWBG region.
We obtain $\Lambda\lesssim10^4$ GeV in Case (i) and $\Lambda\lesssim10^{14}$ GeV in Case (ii). 

Finally, our benchmark points are summarized in Table \ref{tab:bm_summary}.
S1 and S2 have been already discussed in this section. 
S3 and S4 are included as a reference.
In S3, although the strong first-order EWPT is possible, no significant deviations appear in the Higgs coupling constants
since $\alpha=0$.
Such a specific case was pointed out in Ref.~\cite{Ashoorioon:2009nf}, and may be probed by  gravitational waves as discussed there.
In S4, the H-S mixing parameter is $\mu_{HS}$. 
Unlike S1, $\alpha$ is not necessarily large in order to induce the strong first-order EWPT.
However, $\lambda_S$ is so large that $\Lambda$ cannot go beyond 10 TeV. 
The low $\Lambda$ is the common feature of the one H-S mixing parameter scenarios, as mentioned above.

\begin{table}[t]
\center
\begin{tabular}{|c|cccc|}
\hline
 & S1 & S2 & S3 & S4 \\
\hline
H-S mixing parameters  & $\lambda_{HS}$ & $\lambda_{HS}, \mu_{HS}$ & $\lambda_{HS},\mu_{HS}$ & $\mu_{HS}$ \\
PT type & D & B & B & B \\
$m_{H_2}$ [GeV] & 500  & 170 & 148 & 500 \\
$\alpha$ [degrees] & 38 & $-20$ & 0 & 20    \\
$v_S$ [GeV] & 200 & 90  & 100 & 200   \\
$\mu_{HS}$ [GeV] & 0.00  & $-80.00$ & $-80.00$  & -310.72 \\
$\mu_S'$ [GeV] & 0 & $-30$ & $-30$ & 0  \\
\hline\hline
$\lambda_H$ & 0.82 & 0.13 & 0.12 & 0.34 \\
$\lambda_S$ & 1.83  & 0.23 & 0.06 & 1.96 \\
$\lambda_{HS}$ & $-2.35$ & 1.08 & 0.80 & 0.00 \\
\hline\hline
$\kappa$ & 0.79 & 0.94 & 1.0 & 0.94 \\
$\Delta\lambda_{H_1H_1H_1}$ [\%] & $-23.7$ & 31.8 & 0.58 & 41.1 \\
$\log_{10}(\Lambda/{\rm GeV})$ & 3.90 & 9.68 & 13.78 & 3.90 \\
\hline\hline
$v_C/T_C$ & $\frac{172.83}{148.87}=1.16$ & $\frac{206.75}{111.76}=1.85$ & $\frac{234.78}{79.31}=2.96$ & $\frac{193.40}{120.53}=$1.60 \\
$v_{SC}$ [GeV] & 145.72 & 98.31 & 100.06  & 182.26 \\
$v_{SC}^{\rm sym}$ [GeV] & 0.00 & 222.33 & 436.99 & 135.40  \\
$\mathcal{E}(T_C)$ & 1.92 & 1.89 & 1.91 & 1.84 \\
$\zeta_{\rm sph}(T_C)$ & 1.14 & 1.18  & 1.18 & 1.20 \\
\hline
\end{tabular}
\caption{The benchmark points for the strong first-order EWPT. $(\lambda_H, \lambda_S, \lambda_{HS})$ are outputs in S1--S3, and $(\lambda_H, \lambda_S, \mu_{HS})$ are outputs in S4. 
$\mu_S=0$ is taken throughout our analysis. 
For details, see the text. S1 is already disfavored by the LHC data and EW precision tests.}
\label{tab:bm_summary}
\end{table}

\section{Conclusions and Discussion}\label{sec:conclusion}
We have reanalyzed the feasibility of the strong first-order EWPT in the rSM.
The sphaleron decoupling condition is improved by taking account of the one-loop corrections 
at zero and nonzero temperatures. 
As explicitly shown in this paper, the sphaleron energy gets lowered at high temperatures, 
and thus the sphaleron decoupling condition becomes more severe. 
For moderate values of the model parameters, the sphaleron decoupling condition
is found to be $v_C/T_C> (1.1-1.2)$,
which is 10$-$20\% more stringent than the conventional one.

We also investigated the impacts of the improved sphaleron decoupling condition on the deviations 
of the Higgs coupling constants from the SM values.
In a typical case, if the Higgs couplings to the gauge bosons/fermions deviate from the SM values by about 3 (10)\%, 
the deviation of the triple Higgs boson coupling can be as large as about 16 (50)\%, 
which is about 4 (8)\% larger than that based on the conventional criterion $v_C/T_C>1$.
It is also found that the ranges of $m_{H_2}$ that are consistent with the successful sphaleron decoupling
get limited by certain amounts 
depending on the magnitude of $\kappa$. For $\kappa\simeq 0.90$, we observe that the upper bound of 
$m_{H_2}$ is reduced by about 20 GeV if the refined sphaleron decoupling condition is used.

Apart from the $\alpha=0$ case, as in S3, significant deviations from the SM values show up 
in the Higgs coupling constants if the EWPT is strongly first order. 
Such deviations can be probed at the high-luminosity LHC~\cite{HL-LHC}  
and the International Linear Collider~\cite{ILC}. 

Finally, we comment on some remaining issues.
In order to reduce the theoretical uncertainties in the sphaleron decoupling condition, 
we should include the subleading contributions omitted here.
For example, the translational and rotational zero-mode factors around the sphaleron 
can have some effects, leading to the enhanced $\zeta_{\rm sph}$.
In addition, $T_C$ has to be replaced with $T_N$ in the sphaleron decoupling condition.
Since some regions in the parameter space show a significantly large $v_C/T_C$, 
the corresponding supercooling can be sizable, which delays the onset of the EWPT. 
If the EWPT mostly proceeds via bubble nucleation rather than bubble expansion, 
the EWBG mechanism might not work. If this is the case, we may obtain the upper bound of the size of $v_C/T_C$, which would further limit the feasible regions. 

In order to incorporate new $CP$ violation, the current model has to be extended.
As long as new particles do not affect the EWPT significantly, the current analysis would be valid. 

\appendix

\begin{acknowledgments}
We thank Junji Hisano and Natsumi Nagata for valuable comments.
\end{acknowledgments}



\begin{thebibliography}{99}
\bibitem{Aad:2012tfa} 
  G.~Aad {\it et al.}  [ATLAS Collaboration],
  Phys.\ Lett.\ B {\bf 716}, 1 (2012)
  [arXiv:1207.7214 [hep-ex]].
  
\bibitem{Chatrchyan:2012ufa} 
  S.~Chatrchyan {\it et al.}  [CMS Collaboration],
  Phys.\ Lett.\ B {\bf 716}, 30 (2012)
  [arXiv:1207.7235 [hep-ex]].

\bibitem{Beringer:1900zz} 
  J.~Beringer {\it et al.}  [Particle Data Group Collaboration],
  Phys.\ Rev.\ D {\bf 86}, 010001 (2012).

\bibitem{Sakharov:1967dj}
  A.~D.~Sakharov,
  Pisma Zh.\ Eksp.\ Teor.\ Fiz.\  {\bf 5}, 32 (1967)
  [JETP Lett.\  {\bf 5}, 24 (1967\ SOPUA,34,392-393.1991\ UFNAA,161,61-64.1991)].

\bibitem{sm_ewpt}
  K.~Kajantie, M.~Laine, K.~Rummukainen and M.~E.~Shaposhnikov,
  Phys.\ Rev.\ Lett.\  {\bf 77}, 2887 (1996);~
%
  K.~Rummukainen, M.~Tsypin, K.~Kajantie, M.~Laine and M.~E.~Shaposhnikov,
  Nucl.\ Phys.\  B {\bf 532}, 283 (1998);~
%
  F.~Csikor, Z.~Fodor and J.~Heitger,
  Phys.\ Rev.\ Lett.\  {\bf 82}, 21 (1999);~
%
  Y.~Aoki, F.~Csikor, Z.~Fodor and A.~Ukawa,
  Phys.\ Rev.\  D {\bf 60}, 013001 (1999).

\bibitem{Cabibbo:1963yz} 
  N.~Cabibbo,
  Phys.\ Rev.\ Lett.\  {\bf 10}, 531 (1963).

\bibitem{Kobayashi:1973fv} 
  M.~Kobayashi and T.~Maskawa,
  Prog.\ Theor.\ Phys.\  {\bf 49}, 652 (1973).

\bibitem{ewbg_sm_cp}
  M.~B.~Gavela, P.~Hernandez, J.~Orloff and O.~Pene,
  Mod.\ Phys.\ Lett.\  A {\bf 9} (1994) 795;~
%
  M.~B.~Gavela, P.~Hernandez, J.~Orloff, O.~Pene and C.~Quimbay,
  Nucl.\ Phys.\  B {\bf 430} (1994) 382;~
%
  P.~Huet and E.~Sather,
  Phys.\ Rev.\  D {\bf 51} (1995) 379;~
%
  T.~Konstandin, T.~Prokopec and M.~G.~Schmidt,
  Nucl.\ Phys.\  B {\bf 679} (2004) 246.

\bibitem{ewbg}
  V.~A.~Kuzmin, V.~A.~Rubakov and M.~E.~Shaposhnikov,
  Phys.\ Lett.\ B {\bf 155} (1985) 36.
For reviews on electroweak baryogenesis, see
A.~G.~Cohen, D.~B.~Kaplan and A.~E.~Nelson,
Ann.\ Rev.\ Nucl.\ Part.\ Sci.\  {\bf 43} (1993) 27;
%
M.~Quiros,
Helv.\ Phys.\ Acta {\bf 67} (1994) 451;
%
V.~A.~Rubakov and M.~E.~Shaposhnikov,
Usp.\ Fiz.\ Nauk {\bf 166} (1996) 493;
%
K.~Funakubo,
Prog.\ Theor.\ Phys.\  {\bf 96} (1996) 475;
%
M.~Trodden,
Rev.\ Mod.\ Phys.\  {\bf 71} (1999) 1463;
%
W.~Bernreuther,
Lect.\ Notes Phys.\  {\bf 591} (2002) 237;
%
  J.~M.~Cline,
  [arXiv:hep-ph/0609145];
%
  D.~E.~Morrissey and M.~J.~Ramsey-Musolf,
  New J.\ Phys.\  {\bf 14}, 125003 (2012);~
%
  T.~Konstandin,
  arXiv:1302.6713 [hep-ph].

\bibitem{MSSM-EWBG_LHCtension}  
  T.~Cohen, D.~E.~Morrissey and A.~Pierce,
  Phys.\ Rev.\ D {\bf 86}, 013009 (2012);~
%
  D.~Curtin, P.~Jaiswal and P.~Meade,
  JHEP {\bf 1208}, 005 (2012);~
%
  M.~Carena, G.~Nardini, M.~Quiros and C.~E.~M.~Wagner,
  JHEP {\bf 1302}, 001 (2013);~
 %
  K.~Krizka, A.~Kumar and D.~E.~Morrissey,
  arXiv:1212.4856 [hep-ph].
%

\bibitem{Funakubo:2009eg} 
  K.~Funakubo and E.~Senaha,
  Phys.\ Rev.\ D {\bf 79}, 115024 (2009)
  [arXiv:0905.2022 [hep-ph]].

\bibitem{EWBG_rSM}
  M.~Dine, P.~Huet, R.~L.~Singleton, Jr and L.~Susskind,
  Phys.\ Lett.\ B {\bf 257}, 351 (1991);~
  %
  Y.~Kondo, I.~Umemura and K.~Yamamoto,
  Phys.\ Lett.\ B {\bf 263}, 93 (1991);~
%
  J.~Choi and R.~R.~Volkas,
  Phys.\ Lett.\ B {\bf 317}, 385 (1993)
  [hep-ph/9308234];~
%
  J.~R.~Espinosa and M.~Quiros,
  Phys.\ Lett.\ B {\bf 305}, 98 (1993)
  [hep-ph/9301285];~
%
  K.~E.~C.~Benson,
  Phys.\ Rev.\ D {\bf 48}, 2456 (1993);~
%
  J.~McDonald,
  Phys.\ Lett.\ B {\bf 323}, 339 (1994);~
%
  G.~C.~Branco, D.~Delepine, D.~Emmanuel-Costa and F.~R.~Gonzalez,
  Phys.\ Lett.\ B {\bf 442}, 229 (1998)
  [hep-ph/9805302];~
%
  S.~W.~Ham, Y.~S.~Jeong and S.~K.~Oh,
  J.\ Phys.\ G {\bf 31}, 857 (2005)
  [hep-ph/0411352];~
  J.~M.~Cline, G.~Laporte, H.~Yamashita and S.~Kraml,
  JHEP {\bf 0907}, 040 (2009)
  [arXiv:0905.2559 [hep-ph]];~
  %
  S.~Das, P.~J.~Fox, A.~Kumar and N.~Weiner,
  JHEP {\bf 1011}, 108 (2010)
  [arXiv:0910.1262 [hep-ph]];~
  %
  D.~J.~H.~Chung and A.~J.~Long,
  Phys.\ Rev.\ D {\bf 84}, 103513 (2011)
  [arXiv:1108.5193 [astro-ph.CO]];~
%
  J.~R.~Espinosa, B.~Gripaios, T.~Konstandin and F.~Riva,
  JCAP {\bf 1201}, 012 (2012)
  [arXiv:1110.2876 [hep-ph]];~
%
  M.~Fairbairn and R.~Hogan,
  JHEP {\bf 1309}, 022 (2013)
  [arXiv:1305.3452 [hep-ph]];~
  P.~H.~Damgaard, D.~O'Connell, T.~C.~Petersen and A.~Tranberg,
  Phys.\ Rev.\ Lett.\  {\bf 111}, no. 22, 221804 (2013)
  [arXiv:1305.4362 [hep-ph]];~
%
  T.~Li and Y.~-F.~Zhou,
  arXiv:1402.3087 [hep-ph].

\bibitem{Profumo:2007wc} 
  S.~Profumo, M.~J.~Ramsey-Musolf and G.~Shaughnessy,
  JHEP {\bf 0708}, 010 (2007)
  [arXiv:0705.2425 [hep-ph]].

\bibitem{Gonderinger:2009jp} 
  M.~Gonderinger, Y.~Li, H.~Patel and M.~J.~Ramsey-Musolf,
  JHEP {\bf 1001}, 053 (2010)
  [arXiv:0910.3167 [hep-ph]].
%
\bibitem{Ashoorioon:2009nf} 
  A.~Ashoorioon and T.~Konstandin,
  JHEP {\bf 0907}, 086 (2009)
  [arXiv:0904.0353 [hep-ph]].

\bibitem{Espinosa:2011ax} 
  J.~R.~Espinosa, T.~Konstandin and F.~Riva,
  Nucl.\ Phys.\ B {\bf 854}, 592 (2012)
  [arXiv:1107.5441 [hep-ph]].

\bibitem{Cline:2012hg} 
  J.~M.~Cline and K.~Kainulainen,
  JCAP {\bf 1301}, 012 (2013)
  [arXiv:1210.4196 [hep-ph]].

\bibitem{Kanemura:2004ch} 
  S.~Kanemura, Y.~Okada and E.~Senaha,
  Phys.\ Lett.\ B {\bf 606}, 361 (2005)
  [hep-ph/0411354].


\bibitem{Funakubo:2005pu} 
  K.~Funakubo, S.~Tao and F.~Toyoda,
  Prog.\ Theor.\ Phys.\  {\bf 114}, 369 (2005)
  [hep-ph/0501052].
  
\bibitem{Coleman:1973jx} 
  S.~R.~Coleman and E.~J.~Weinberg,
  Phys.\ Rev.\ D {\bf 7}, 1888 (1973).
  
\bibitem{Jackiw:1974cv} 
  R.~Jackiw,
  Phys.\ Rev.\ D {\bf 9}, 1686 (1974).
  
\bibitem{Dolan:1973qd} 
  L.~Dolan and R.~Jackiw,
  Phys.\ Rev.\ D {\bf 9}, 3320 (1974).

\bibitem{Carrington:1991hz} 
  M.~E.~Carrington,
  Phys.\ Rev.\ D {\bf 45}, 2933 (1992).
   
\bibitem{2stepPT}    
  D.~Land and E.~D.~Carlson,
  Phys.\ Lett.\ B {\bf 292}, 107 (1992)
  [hep-ph/9208227];~
 %
  A.~Hammerschmitt, J.~Kripfganz and M.~G.~Schmidt,
  Z.\ Phys.\ C {\bf 64}, 105 (1994)
  [hep-ph/9404272].
     

\bibitem{Arnold:1987mh} 
  P.~B.~Arnold and L.~D.~McLerran,
  Phys.\ Rev.\ D {\bf 36}, 581 (1987).

\bibitem{Hellmund:1991ub} 
  M.~Hellmund and J.~Kripfganz,
  Nucl.\ Phys.\ B {\bf 373}, 749 (1992).
  
\bibitem{D'Onofrio:2012jk} 
  M.~D'Onofrio, K.~Rummukainen and A.~Tranberg,
  JHEP {\bf 1208}, 123 (2012)
  [arXiv:1207.0685 [hep-ph]].
    
\bibitem{Manton:1983nd} 
  N.~S.~Manton,
  Phys.\ Rev.\ D {\bf 28}, 2019 (1983).

\bibitem{Klinkhamer:1984di} 
  F.~R.~Klinkhamer and N.~S.~Manton,
  Phys.\ Rev.\ D {\bf 30}, 2212 (1984).

\bibitem{Funakubo:2005bu}
  K.~Funakubo, A.~Kakuto, S.~Tao and F.~Toyoda,
  Prog.\ Theor.\ Phys.\  {\bf 114}, 1069 (2006)
  [arXiv:hep-ph/0506156].

\bibitem{sph_rSM}
  B.~M.~Kastening and X.~Zhang,
  Phys.\ Rev.\ D {\bf 45}, 3884 (1992);~
  %
  J.~Choi,
  Phys.\ Lett.\ B {\bf 345}, 253 (1995)
  [hep-ph/9409360];~
  A.~Ahriche,
  Phys.\ Rev.\ D {\bf 75}, 083522 (2007)
  [hep-ph/0701192].
  %
  %

\bibitem{sph_wU1Y}
  B.~Kleihaus, J.~Kunz and Y.~Brihaye,
  Phys.\ Lett.\ B {\bf 273}, 100 (1991);~
%
  F.~R.~Klinkhamer and R.~Laterveer,
  Z.\ Phys.\ C {\bf 53}, 247 (1992).


\bibitem{Zeldovich:1974uw} 
  Y.~B.~Zeldovich, I.~Y.~.Kobzarev and L.~B.~Okun,
  Zh.\ Eksp.\ Teor.\ Fiz.\  {\bf 67}, 3 (1974)
  [Sov.\ Phys.\ JETP {\bf 40}, 1 (1974)].
  
\bibitem{Peskin:1990zt} 
  M.~E.~Peskin and T.~Takeuchi,
  Phys.\ Rev.\ Lett.\  {\bf 65}, 964 (1990).
  
\bibitem{Maksymyk:1993zm} 
  I.~Maksymyk, C.~P.~Burgess and D.~London,
  Phys.\ Rev.\ D {\bf 50}, 529 (1994)
  [hep-ph/9306267].

\bibitem{Baek:2012uj} 
  S.~Baek, P.~Ko, W.~-I.~Park and E.~Senaha,
  JHEP {\bf 1211}, 116 (2012)
  [arXiv:1209.4163 [hep-ph]].

\bibitem{ATLAS}
 ATLAS collaboration, Updated coupling measurements of the Higgs boson with the ATLAS detector using up to 25 fb$^{-1}$ of proton-proton collision data, ATLAS-CONF-2014-009.

\bibitem{CMS}
 CMS collaboration, Combination of standard model Higgs boson searched and measurements of the properties of the new boson with a mass near 125 GeV, CMS-PAS-HIG-13-004.

\bibitem{Noble:2007kk} 
  A.~Noble and M.~Perelstein,
  Phys.\ Rev.\ D {\bf 78}, 063518 (2008)
  [arXiv:0711.3018 [hep-ph]].

\bibitem{Ahriche:2013vqa} 
  A.~Ahriche, A.~Arhrib and S.~Nasri,
  JHEP {\bf 1402}, 042 (2014)
  [arXiv:1309.5615 [hep-ph]].
  
\bibitem{lamhhh}
  S.~Kanemura, S.~Kiyoura, Y.~Okada, E.~Senaha and C.~P.~Yuan,
  Phys.\ Lett.\ B {\bf 558}, 157 (2003);~
  %
  S.~Kanemura, Y.~Okada, E.~Senaha and C.~-P.~Yuan,
  Phys.\ Rev.\ D {\bf 70}, 115002 (2004)
  [hep-ph/0408364].

\bibitem{Kanemura:2012hr} 
  S.~Kanemura, E.~Senaha, T.~Shindou and T.~Yamada,
  JHEP {\bf 1305}, 066 (2013)
  [arXiv:1211.5883 [hep-ph]].


\bibitem{HL-LHC}
  [ ATLAS Collaboration],
  ``Physics at a High-Luminosity LHC with ATLAS,''
  arXiv:1307.7292 [hep-ex].

\bibitem{ILC}
%
  H.~Baer, T.~Barklow, K.~Fujii, Y.~Gao, A.~Hoang, S.~Kanemura, J.~List and H.~E.~Logan {\it et al.},
  arXiv:1306.6352 [hep-ph].

\end{thebibliography}
\end{document}